

\input harvmac.tex

\def\np{Nucl. Phys.}
\def\pl{Phys. Lett.}

\def\cmp{ Comm. Math. Phys.}
\def\ijmp{Int. J. Mod. Phys.}

\def\pr{Phys. Rep.}
\def\Ph{{\cal M}}

\def\Phai{{\cal M}_i}
\def\Phc{{\cal M}_1}
\def\Phn{{\cal M}_2}
\def\Phl{{\cal M}_3}
\def\Ns{{\cal N}}
\def\Phcs{\Phc/\Ns}
\def\Phns{\Phn/\Ns}

\def\G{$SL(2,R)$}

\def\wfN{\Psi_N (\tau;z)}
\def\Hc{ {\bf \cal H}_{\Phcs}}
\def\Hn{ {\bf \cal H}_{\Phns}}

\def\Hcm{ {\bf \cal H}_{\Phcs}^{-}}
\def\Hcp{ {\bf \cal H}_{\Phcs}^{+}}
\def\HCS{ {\bf \cal H}_{CS}}

\def\tHp{{\cal O}_{p/q}}
\def\tHq{{\cal O}_{q/p}}
\def\tHpq{{\cal O}_{pq}}
\def\eigen{\lambda_n^{(m)}}
\def\vopa{\Phi_n(a)}
\def\vopb{\Phi_n(b)}

\overfullrule=0pt

{\nopagenumbers
\font\bigrm=cmb10 scaled\magstep1
\rightline{GEF-TH 2/1993}
\ \medskip
\centerline{\bigrm NEW MODULAR REPRESENTATIONS AND}

\centerline{\bigrm FUSION ALGEBRAS FROM QUANTIZED}

\centerline{\bigrm SL(2,R) CHERN-SIMONS THEORIES}
\vskip 1truecm
\centerline{CAMILLO IMBIMBO}
\vskip 5pt
\centerline{\it I.N.F.N., Sezione di Genova}
\centerline{\it Via Dodecaneso 33, I-16146 Genova, Italy}
\vskip 8pt
\vskip 2truecm
\centerline{ABSTRACT}

We consider the quantum-mechanical algebra of observables
generated by canonical quantization of \G\ Chern-Simons
theory with rational charge on a space manifold with torus topology.
We produce modular representations generalizing the representations
associated to the $SU(2)$ WZW models and we exhibit the explicit
polynomial representations of the corresponding fusion algebras.
The relation to Kac-Wakimoto characters
of highest weight $\widehat{sl}(2)$ representations with
rational level is illustrated.

\ \vfill

\leftline{GEF-TH 2/1993}
\leftline{January 1993}
\eject     }

\beginsection 1. INTRODUCTION

\indent Three-dimensional topological Chern-Simons gauge theory with gauge
group $G$ is described by the following action
\ref\witjones{E. Witten,\cmp\ {\bf 121} (1988) 351.}:
\eqn\csaction {S = {k\over 2\pi}\int_{M_3}
\eta_{ab} A^a dA^b + {2\over 3} f_{abc} A^a A^b A^c}
where $f_{abc}$ are the structure constants of the
Lie algebra of $G$, $\eta_{ab}$ is the invariant Killing metric
on it, and $A^a$ are the gauge field 1-forms.

Chern-Simons theory with $G$ compact was solved non-pertubatively
by means of ``holomorphic'' canonical quantization methods
which uncovered its relation to the two-dimensional Wess-Zumino-Witten
model on the group manifold $G$ \ref\elitzur{S. Elitzur,
G. Moore, A. Schwimmer and N. Seiberg, \np\ {\bf B326} (1989) 108.}%
\nref\bos{M. Bos and V.P. Nair, \pl\ {\bf B223} (1989)
61; \ijmp\ {\bf A5} (1990) 959.}-\ref\axelrod{S. Axelrod, S. Della
Pietra and E. Witten, J. Diff. Geom. {\bf 33} (1991) 787.}.
In this talk we will discuss the much less well understood
Chern-Simons theory with non-compact
gauge group $G=SL(2,R)$ \ref\hverlinde{H. Verlinde, Princeton preprint,
PUTP-89/1140, unpublished.}%
\nref\ehverlinde{E. Verlinde and H. Verlinde, Princeton preprint,
PUTP-89/1149, unpublished.}%
\nref\trimb{C. Imbimbo, {\it in}: ``String Theory and Quantum Gravity
`91,'' H. Verlinde, ed., World Scientific, Singapore (1992).}-\ref\cimbimbo
{C. Imbimbo, \np\ {\bf B384} (1992) 484.}. Since the meaning of the \G\ WZW
model
as a conformal quantum field theory is far from clear \ref\furlan{P. Furlan,
R. Paunov, A.Ch. Ganchev and V.B. Petkova, \pl\ {\bf 276} (1991)
63; A.Ch. Ganchev and V.B. Petkova, Trieste preprint, SISSA-111/92/EP.},
the relation of \G\ Chern-Simons theory to two-dimensional conformal
field theory, if it exists, must be of a novel type.
In what follows we will describe various features (such as modular
properties and fusion algebras) of the yet unknown two-dimensional
counterpart of \G\ Chern-Simons theory.

Let us first briefly review the main aspects of the
holomorphic canonical quantization of the action
\csaction\ to point out why it does not extend to theories
with non-compact gauge groups.
In the Hamiltonian formalism, the three-dimensional space-time
manifold $M_3$ is
the product $\Sigma \times R^1$ of a two-dimensional compact
surface $\Sigma$ and of the time axis $R^1$. Going to the $A_0=0$ gauge,
one obtains a free gauge-fixed action
\eqn\free{S ={ k\over 2\pi}\int dt \int_{\Sigma} \epsilon^{ij}
\eta_{ab} A_i^a {\dot A}_j^b d^2x}
\noindent where $\epsilon^{ij}$ is the anti-symmetric tensor on the
two-dimensional space manifold $\Sigma$. The constraint
\eqn\gauss{ \epsilon^{ij}F^a_{ij} =0}
\noindent associated to the gauge-fixing encodes the non-linearity
of the theory. The ``Gauss law'' \gauss\
states that the classical {\it physical} phase space $\Ph$ is the
space of flat G-connections on the two-dimensional surface $\Sigma$.

In the holomorphic quantization, one selects a complex structure on
$\Sigma$ which determines a complex structure on the space of G-connections
on $\Sigma$. States, in the ``quantize-first'' approach, are then described
by holomorphic wave-functionals $\Psi (A_z^a)$ which depend only
on the holomorphic components of the gauge field 1-forms
$A^a = A_z^a dz + A_{\bar z}^a d{\bar z}$ and which are normalizable
in the scalar product
\eqn\scalar{ < \Psi_1 , \Psi_2> = \int [DA_z DA_{\bar z}]
e^{- {k \over 2\pi}\int_{\Sigma}\eta_{ab} A^a_z A^b_{\bar z}}
{\bar \Psi_1}(A_{\bar z}) \Psi_2 (A_z). }

Physical states $\Psi (A_z^a)$ are normalizable solutions of the functional
equation which is the quantum version of the
``Gauss law'' constraint \gauss :
\eqn\ward{ D_z^{ab} {\delta \over \delta A_z^b} \Psi = {k \over 2\pi}
{\bar \partial}A^a_z \Psi .}

Eqs.\ward\ are identical to the Ward Identities for the generating
functional of current correlators of the two-dimensional
WZW model on the Riemann surface $\Sigma$.
This shows that for $G$ compact,
the vector space of physical states of the Chern-Simons theory \csaction\
is isomorphic to the space of current blocks of the
two-dimensional ${\hat G}$ current algebra. The central task
of canonical quantization of Chern-Simons theory is to
prove that the unitary structure \scalar\ on the vector space
of quantum states of the Chern-Simons theory is in fact the
same as the natural unitary structure on the current blocks
of the WZW theory for which modular transformations are represented
by unitary matrices. For $G$ compact, this has been proven
explicitly for $\Sigma$ of genus zero and one
\axelrod ,\ref\imbimbo{C. Imbimbo, \pl\ {\bf B258} (1991) 353.}.

When $G$ is non-compact, $\eta_{ab}$ is not positive-definite,
and the scalar product \scalar\ is not, even formally, well
defined. In this case, the holomorphic gauge-invariant
polarization $\Psi (A_z^a)$ does not define a genuine
positive-definite K\"ahler structure on the space of $G$ connections
on $\Sigma$. If $G$ is a non-compact but complex group,
one can find a family of {\it real} gauge-invariant polarizations
which is preserved by the reparametrizations of $\Sigma$
\ref\witcomplex{E. Witten, \cmp\ {\bf 137} (1991) 29.}.
The existence of a reparametrization invariant family
of gauge-invariant polarizations is the essential
condition that makes it possible to discuss topological invariance
at quantum level.

When $G=SL(2,R)$ a reparametrization
invariant family of gauge-invariant (positive-definite) polarizations does
not exist \witcomplex . This is the main difficulty
in quantizing the theory in a topological invariant way, or, equivalently,
in proving that the mapping class group is implemented unitarily
on the Hilbert space of quantum states.
The difficulty is analogous to the one that
is met when quantizing the Heisenberg algebra $[x_\mu, p_\nu]=
i\eta_{\mu \nu}$ with a Lorentzian type of metric $\eta_{\mu \nu}$.
Choosing a real polarization, one obtains states represented
by wave-functions $\psi(x_{\mu})$, and $p_{\mu} = -i{\partial}_{\mu}$.
This gives a perfectly unitary (but not highest-weight) representation
of the Heisenberg algebra. This representation is, however,
{\it not} equivalent to the (heighest-weight but not
positive-definite) Fock representation
defined by the creation and annihilation operators
$a_{\mu}^\dagger = x_{\mu} + ip_{\mu}, a_{\mu}= x_{\mu} - ip_{\mu}$.
The point is that though dependence on the polarization is a `fact of life'
of the quantization process, it potentially jeopardizes quantum
topological invariance of the Chern-Simons theory.

Because of this fundamental difficulty which affects
any attempt  to ``first-quantize'' the full space of two-dimensional
\G\ connections on $\Sigma$ and to impose the ``Gauss-law''
\gauss\ as an operatorial constraint on the physical states,
we will work in the so-called ``constrain-first'' approach
\elitzur ,\imbimbo\ in which one quantizes directly  the {\it physical}
classical phase space $\Ph$. Since $\Ph$ is finite-dimensional,
the canonical quantization problem actually has a finite number of
degrees of freedom.  However, the topology of $\Ph$ is, for a
generic $\Sigma$, quite intricate. For this reason, we will
restrict ourselves to the case when $\Sigma$ has the torus
topology; such a limitation has been sufficient to unravel the
underlying two-dimensional current algebra structure in the
case of the compact gauge group $SU(2)$ \imbimbo .

When $\Sigma$ is a torus, the problem of quantizing $\Ph$ is
reduced to the problem of quantizing the moduli space of flat-connections
of an {\it abelian} gauge group \axelrod . This makes the
computation for genus one drastically simpler than
for higher genus, where non-abelian Chern-Simons theory appears to be
vastly more complex than abelian.
On the other hand, the factorization properties of 2-dimensional
conformal field theories suggest that
the torus topology already contains most, if not all, of the
complexities of higher genus. The solution of this apparent paradox
is that $\Ph$ for a torus is {\it almost}  the space of flat
connections of an abelian group, but not quite: it is the space of
abelian flat connections modulo the action of a discrete group whose fixed
points give rise to orbifold singularities. It is only here that
the quantization of non-abelian Chern-Simons theory with compact
gauge group for genus one differs from the computationally trivial abelian
case. The projection associated to such a discrete group
is responsible for the emergence of a non-abelian
structure for the \G\ Chern-Simons theory as well.

\beginsection 2. QUANTIZATION  OF $\Ph$

\indent The difficulties of the ``quantize-first'' approach for
\G\ gauge group have their counterpart in the ``constrain-first''
method in the singular geometry of $\Ph$. Corresponding
to the three types of inequivalent Cartant subgroups of \G ,
there exist three different ``branches''
of $\Ph$ which have a non-vanishing intersection:
$\Ph = \bigcup_{i=1,2,3} \Phai,$ with $i=1,2,3$.
It is interesting to notice that similarly ``branched'' phase
spaces appear also in the context of two-dimensional topological
theories based on the gauged \G\ WZW model and related to solvable string
theories \ref\witmatrix{E. Witten, \np\ {\bf B371} (1991) 191.}.

Flat \G\ connections on a torus correspond to pairs $(g_1,g_2)$
of commuting
\G\ elements modulo overall conjugation in \G .
The elements $g_1$ and $g_2$
represent the holonomies of the flat connections around the two
non-trivial cycles of the torus.

The $\Phc$ branch is made of flat connections whose holonomies can
be simultaneously brought by conjugation into the compact $U(1)$
subgroup of \G . Therefore, $\Phc \approx T^{(1)}$, the two
dimensional torus.

$\Phn$ is the branch of flat connections whose holonomies,
when represented by $2\times 2$ real unimodular matrices,
can be conjugated into a diagonal form: i.e.
$g_i = \left(\matrix{\pm e^{x_i}&0\cr 0&\pm e^{-x_i}}\right),$ with
$i=1,2$. However, one can still conjugate diagonal holonomies by an element of
the gauge group which permutes the eigenvalues, mapping
$x_i$ onto $-x_i$. Therefore, $\Phn$ consists
of four copies of $R^{(2)}/Z_2$ whose origins are attached to the
four points of $\Phc$ which correspond to flat connections with
holonomies in the center of the gauge group \G .

Finally, $\Phl$ is the branch of flat connections
with holonomies which can be conjugated into an upper triangular
form with units on the diagonal. Conjugation allows one to rescale the
(non-vanishing) elements in the upper right corner by an arbitrary positive
number. Thus, $\Phl \approx S^1$, the real circle.
Being odd-dimensional, $S^1$ cannot be a
genuine non-degenerate symplectic space. In fact,
when pushed down to $\Phl$, the symplectic form
on the space of flat connections coming from the Chern-Simons action
vanishes identically. $\Phl$ represents
a ``null'' direction for the symplectic form of the \G\ Chern-Simons
theory, reflecting the indefiniteness of the \G\ Killing form.
Since $\Phl$ is a {\it disconnected} piece of the total phase space $\Ph$,
it is consistent to consider the problem of quantizing $\Phc \cup \Phn$
independently of $\Phl$.

There are no ``rigorous'' ways to quantize a phase space
consisting of different branches with a non-zero intersection.
The strategy adopted both in \cimbimbo\ and \witmatrix\
is to consider the smooth, non-compact manifold $\Ph /\Ns$
obtained by deleting the intersection $\Ns \equiv \Phc \cap
\Phn$ of the two branches of $\Ph$. $\Ph /\Ns$ consists of
disconnected smooth components $\Phc /\Ns$ and $\Phn /\Ns$,
which, upon quantization, give rise to Hilbert spaces
of wave functions $\Hc$ and $\Hn$.
It seems reasonable to think of a wave function on the union
$\Phc \cup \Phn$ as a pair $(\psi_1,\psi_2)$ of wave functions,
with $\psi_1 \in \Hc$ and $\psi_2 \in \Hn$, ``agreeing''
in some sense on the intersection $\Ns$. The proposal
of \cimbimbo\ is that $\psi_1$ and $\psi_2$, when represented by
holomorphic functions, should have the same
behaviour around the points in $\Ns$.
This implies that the pair $(\psi_1,\psi_2)$ should be determined
uniquely by $\psi_1$ and
that most of the states $\psi_2$ in the infinite-dimensional
$\Hn$ should be discarded.
The conclusion of the analysis in \cimbimbo\ is that
the quantization of $\Ph$ produces a Hilbert space $\HCS$
which is a subspace of $\Hc$ with definite parity under
the conjugation operator $C$,
\eqn\reflections{C : (g_1,g_2)\rightarrow (g_1^{-1}, g_2^{-1}).}

The space $\Hc$ coming from the quantization of $\Phcs$ is
the representation space of the 't Hooft algebra
\ref\tHft{G. 't Hooft, \np\ {\bf B138} (1978) 1.}:
\eqn\tHooft{ AB = \mu BA ,}
\noindent where $\mu$ is a phase related to the coupling constant
appearing in the \G\ Chern-Simons action \csaction\ through
the equation:
\eqn\cocycle{ \mu =e^{{i\pi \over k}}.}
The quantum operators $A$ and $B$ are the quantum versions of
the classical holonomies $g_1$ and $g_2$.
Unitary, irreducible representations of \tHooft\ are finite-dimensional
when $k$ is {\it rational}:
\eqn\charge{2k= 2s/r = p/q ,}
\noindent with $s,r$ and $p,q$ coprime integers.

Since we are interested in investigating the connection between \G\
Chern-Simons
theory and two-dimensional {\it rational} conformal field theories,
we will restrict ourselves to $k$ rational
as in Eq.\charge, and we will denote the corresponding 't Hooft
algebra by $\tHp$.

Modular transformations are external automorphisms of the
algebra $\tHp$:
\eqn\auto{S:\cases{A\rightarrow B^{-1}\cr B\rightarrow A\cr}\;\;
T:\cases{A\rightarrow A\cr B\rightarrow \mu^{-1/2}AB\cr}\;\;
C:\cases{A\rightarrow A^{-1}\cr B\rightarrow B^{-1}.\cr}}
One can verify that $S,T,C$ satisfy the modular
group relations, $S^2 = C$ and $(ST)^3=1$ and that the conjugation
operator $C$ commutes with the modular group generators, $SC=CS$, $TC=CT$.

The requirement that the automorphisms $S,T,C$ be represented
unitarily on the re\-presentation space $\Hc$ of the 't Hooft algebra
\tHooft\ selects a unique (up to equivalence) unitary,
irreducible representation of $\tHp$ with dimension $p$:
\eqn\representation{\eqalign{ (A)_{MN}&=(-1)^{pq} \mu^N \delta_{N,M}
\cr (B)_{MN}& = (-1)^{pq}\delta_{M,N+1},
\;\;\; M,N= 0,1,...,p-1.\cr}}
The corresponding unitary representation of the modular group is:
\eqn\reprqp{\eqalign{(S)_{MN}&=
{1\over \sqrt{p}}e^{2\pi i {q \over p}M N} \cr
(T)_{MN}&= (-1)^{Npq}e^{2\pi i{q \over 2p}N^2 - 2\pi i \theta (q;p)/3}
\delta_{N,M}  \cr
(C)_{MN}&= \delta_{N+M,0}\;\;\; M,N = 0,1,...,p-1, \cr } }
where the phase $\theta (q;p)$ is determined by the $SL(2,Z)$ relation
$(ST)^3=1$ and can be written as a generalized Gauss sum:
\eqn\tetapq{ e^{2\pi i \theta(q;p)} = {1\over \sqrt{p}}\sum_{n=0}^{p-1}
(-1)^{pqn}e^{2\pi i{q \over 2p} n^2}. }
An explicit formula for $\theta (q;p)$ has been found in \cimbimbo .

The modular invariant representations \representation\
of the 't Hooft algebra \tHooft\ admit concrete realizations
in terms of {\it holomorphic} functions only if $2k$ is integer
(i.e., if $q=1$). When $q\not= 1$, a holomorphic realization
of \representation\ involves rather $q$-multiplets of holomorphic
functions
\ref\jengo{R. Jengo and K. Lechner, \pr\ {\bf 213} (1992) 179.},\trimb .
Geometrically this can be understood
by noting that the compact $\Phc$ is quantizable, in the
sense of geometric quantization \ref\woodhouse{
N. Woodhouse, ``Geometric Quantization,'' Oxford University Press,
Oxford (1980).}, only if $2k$ is an integer. However, the
non-compact $\Phcs$ is quantizable for any real $k$ since
holomorphic wave functions might have non-trivial monodromies
around loops surrounding points of $\Ns$. For $2k=p/q$ rational,
monodromies are represented by $q\times q$ matrices, and
we can think of a holomorphic wave function on $\Phcs$ with
non-trivial monodromy around the points of $\Ns$ as a $q$-multiplet
of wave functions holomorphic on $\Phc$ \cimbimbo .

The holomorphic representation of $\tHp$ is better understood
by considering the following isomorphisms of 't Hooft algebras
\eqn\decomposition{\tHpq \approx \tHq \times \tHp,}
with $\tHq$ and $\tHp$ commuting among themselves. In fact,
denoting by $A$ and $B$, ${\tilde A}$ and ${\tilde B}$,
${\hat A}$ and ${\hat B}$, the generators respectively of the
algebras $\tHp$,$\tHq$ and $\tHpq$, the following relations hold:
\eqn\iso{\eqalign{A&= {\hat A}^q ,\;\; B={\hat B}^q \cr
{\tilde A}&= {\hat A}^p ,\;\; {\tilde B}={\hat B}^p \cr
{\hat A}&= {\tilde A}^{\bar m} A^{\bar n},\;\; {\hat B}=
{\tilde B}^{\bar m} B^{\bar n}, \cr}}
where ${\bar m}$ and ${\bar n}$ are integers determined by
the conditions:
\eqn\identity{\eqalign{1&= {\bar m}p + {\bar n}q \cr
0&\leq {\bar m} \leq q-1, \;\; 0\leq {\bar n}\leq p-1. \cr}}

The modular invariant representation of $\tHpq$ can be
realized on holomorphic theta functions. When $pq$ is
{\it even}, the holomorphic realization of \representation\ is:
\eqn\holoeven{ \Psi_{\lambda} (\tau; z) = \theta_{\lambda, pq/2}
(\tau; z),\;\; \lambda = 0,1,...,pq-1,}
where the $\theta_{n,m}(\tau; z)$ ($n$ integer modulo
$2m$) are level $m$ $SU(2)$ theta functions
\ref\kacalgebras {V.G. Kac, ``Infinite Dimensional Lie Algebras,''
Cambridge University Press, Cambridge (1985).}:
$$\theta_{n,m}(\tau;z)\equiv \sum_{j\in Z}e^{2\pi im\tau (j+{n\over 2m})^2
+2\pi imz(j+{n\over 2m})}.$$
If $pq$ is {\it odd}, the holomorphic, modular invariant realization of
\representation\ is instead:
\eqn\holoodd{ \Psi_{\lambda}(\tau;z)=(-1)^{\lambda}\left(
\theta_{2\lambda+pq,2pq}(\tau;z/2)
- \theta_{2\lambda-pq,2pq}(\tau;z/2)\right).}

Because of the algebra decomposition \decomposition , the representations
\holoeven\ and \holoodd\ decompose into $q$ copies of the representation
\representation\ of $\tHp$. Defining the indices $N$ and $\alpha$ through
\eqn\indices{\lambda = qN +p\alpha,\;\;\; 0\leq N \leq p-1,\;\; 0\leq \alpha
\leq q-1,}
one obtains the $q$-components holomorphic representation of $\tHp$:
\eqn\basis{(\wfN)^{\alpha}=
\theta_{qN + p\alpha,pq/2} (\tau; z/q)}
if $pq$ is even, and
\eqn\basisodd{(\wfN)^{\alpha}=
(-1)^{\lambda}\left(\theta_{q(2N+p)+2p\alpha,2pq}(\tau;z/2q)-
\theta_{q(2N-p)+2p\alpha,2pq}(\tau; z/2q)\right)}
if $pq$ is odd.

The algebra decomposition \decomposition\ implies as well that the
group of external automorphisms of $\tHpq$ also factorizes
into two copies of the modular group commuting among themselves
and acting independently on $\tHq$ and $\tHp$. In particular,
the {\it center} \ ${\cal C}_{pq}$ of the group of external automorphisms
factorizes: ${\cal C}_{pq}= {\cal C}_{q/p} \times {\cal C}_{p/q}$.
Thus, the conjugation operator $C_{pq}\in {\cal C}_{pq}$ of the algebra
$\tHpq$,
which in the representation \holoeven ,\holoodd\ acts as follows
\eqn\conjpq{C_{pq} :\lambda \rightarrow -\lambda ,}
satisfies the equation:
\eqn\conj{C_{pq}= C_{q/p}C_{p/q}= C_{p/q}C_{q/p},}
where $C_{p/q}$ and $C_{q/p}$ are the conjugation operators of the algebras
$\tHp$ and $\tHq$ with action given by
\eqn\pqconj{\eqalign{C_{p/q} &: \lambda \rightarrow {\bar \lambda}\equiv
-qN + p\alpha\cr C_{q/p} &: \lambda \rightarrow -{\bar \lambda}.\cr }}

Since both $C_{p/q}$ and $C_{q/p}$ are in the center ${\cal C}_{pq}$,
it is possible to project the holomorphic representation \holoeven ,
(or \holoodd ) onto subrepresentations with definite values of
$C_{p/q}$ and/or $C_{q/p}$, each one carrying a
unitary representation of the modular group.
It is somewhat remarkable that in this way one obtains
the characters of the $A$ and $D$ diagonal series of
integrable representations of $SU(2)$ current algebra,
the Kac-Wakimoto characters of admissible representations of
\G\ current algebra with fractional level, and the
Rocha-Caridi characters of the completely degenerate
representations of the discrete Virasoro series.

{}From the point of view of the quantization of $\Ph$,
the relevant projection is the one onto the subspace $\Hcm$ ($\Hcp$)
of $\Hc$ with $C_{p/q}=-1$ ($C_{p/q}=1$).
For reasons which are still rather mysterious \axelrod ,\imbimbo ,
when $pq$ is even (odd) the projection onto $\Hcp$ ($\Hcm$) does
not lead to characters related to two-dimensional
conformal field theories. Therefore, in what follows, we will
take as Chern-Simons space $\HCS$ the subspace $\Hcm$, if
$pq$ is even, and $\Hcp$, if $pq$ is odd.

The $C_{p/q}$-odd (or even) combinations of the multi-component wave functions
\basis\ spanning $\HCS$ turn out to be (the numerators of) the
Kac-Wakimoto characters $\chi_{j(n,k);m}(z;\tau)$
\ref\vgkac{V.G. Kac and M. Wakimoto,
Proc. Nat. Acad. Sci. {\bf 85} (1988) 4956.},\ref\mp{S. Mukhi
and S. Panda, \np\ {\bf B338} (1990) 263.} of the irreducible, highest weight
representations of $SL(2,R)$ current algebra with
level $m\equiv t/u$ and spin $j(n,k)= 1/2 (n- k(m+2))$, with
$n=1,2,..., 2u+t-1$ and $k=0,1,..., t-1$.

When $p$ is even and $q$ odd, the explicit relation between Kac-Wakimoto
characters and the $C_{p/q}$-odd combinations $\Psi_N^{(-)}(\tau;z)$
of the Chern-Simons multi-component wave functions \basis\ is:
\eqn\kac{|N>\equiv{({\Psi^{(-)}_N})^\alpha \over \Pi (\tau; z)} =
\cases {\chi_{j (N, 2\alpha);m}(\tau;z) & if
$\alpha \leq {q-1\over 2}$ \cr
-\chi_{j (p/2-N, 2\alpha - q);m}(\tau;z) & if
$\alpha \geq {q+1\over 2}$, \cr}}
where $N=1,...,p/2 -1$, and the level $m$ of the current algebra is related
to the Chern-Simons coupling constant $k$ through the equation
\eqn\level{m+2=k.}
$\Pi (\tau ;z)$ is the Kac-Wakimoto denominator,
\eqn\denominator{ \Pi (\tau ;z) = \theta_{1,2} (\tau ,z) -
\theta_{-1,2} (\tau, z),}
which is holomorphic and non-vanishing on $\Phcs$.
Therefore, the wave functions $\Psi^{(-)}_N (\tau; z)$  and the wave functions
$$ \Psi^\prime_N (\tau;z)=
{\Psi^{(-)}_N (\tau; z) \over \Pi ( \tau; z)}$$
\noindent  appearing in \kac, describe equivalent wave functions
on $\Phcs$, related to each other by a K\"ahler
transformation.

If $p$ is odd and $q$ even, Eqs.\kac\ and \level\ are replaced by
\eqn\kacbis{|N>\equiv{({\Psi^{(-)}_N})^\alpha \over \Pi (\tau; z)} =
\cases {\chi_{j (2N,\alpha);m}(\tau;z) & if $\alpha \leq q/2-1$\cr
-\chi_{j (p-2N,\alpha - q/2);m}(\tau;z) & if $\alpha \geq q/2$,\cr }}
with $N=1,...,{p-1\over 2}$ and the level $m$ given by
\eqn\levelbis{m+2=4k.}

Finally, if both $p$ and $q$ are odd, Eq.\levelbis\ is true
and the relation between $C_{p/q}$-even wave functions and characters becomes:
\eqn\kactris{|N>\equiv {(-1)^{\lambda}(\Psi^{(+)}_N)^{\alpha}
\over \Pi (\tau; z)}=
\chi_{j(p+2N,\alpha);m}(\tau;z/2) +\chi_{j(p-2N,\alpha);m}(\tau;z/2),}
with $N=0,...,{p-1\over 2}.$

Eqs.\kac -\kactris\ generalize the relationship between Chern-Simons
theories and two-dimensional current algebra to the case when the
coupling constant $2k=p/q$ is fractional. To each Chern-Simons state
$|N>$ there correspond not just a single current ``block'' as in the
integer ($q=1$) case, but a {\it q-multiplet} of Kac-Wakimoto characters.
For example, when $p$ is even, the holomorphic representation of
the Chern-Simons state $|N>$ is given by the multiplet
$\{\chi_{j(N,0)}, -\chi_{j(p/2-N,1)}, \chi_{j(N,2)},...,\chi_{j(N,q-1)}\}$.

This is not completely surprising. Though the Kac-Wakimoto characters
share several properties of the characters of two-dimensional conformal
field theories (like modular invariance, unicity of the vacuum, etc.),
they cannot possibly come from a conventional conformal field theory
since the associated fusion rules, as computed from the Verlinde formula
\ref\verfusion{ E. Verlinde, \np\ {\bf B300} (1988) 360.}, may be negative.
On the other hand, as we will show in the next section,
there is a well-defined Verlinde algebra, with positive integer fusion rules,
acting on the Chern-Simons states $|N>$. This is compatible with the
relation that we found between the states $|N>$ and the Kac-Wakimoto
characters,
because of the minus signs appearing in Eqs.\kac ,\kacbis ,\kactris .
In fact, Eqs.\kac -\kactris\ suggest that the
objects relevant to the two-dimensional counterpart of
\G\ Chern-Simons theory are the current algebra ``super-characters''
\eqn\kacnewo{ {\tilde \chi}_{j(n,k)} = (-1)^k \chi_{j(n,k)}}
if $u$ is odd, and
\eqn\kacnewe{ {\tilde \chi}_{j(n,k)} = (-1)^{n+1} \chi_{j(n,k)}}
if $u$ is even. The fusion rules for the ${\tilde \chi}_{j(n,k)}$'s
computed from the Verlinde formula are positive, as
we will check in the next section. The fact that
the Kac-Wakimoto fusion rules can be made simultaneously all positive
by the redefinition \kacnewo ,\kacnewe\ seems to indicate that
the two-dimensional theory underlying the \G\ Chern-Simons theory
(and possibly representing a suitable definition of the
\G\ WZW model) assigns non-trivial ``ghost-parities''
($(-1)^k$ or $(-1)^{n+1}$ for $u$ odd or even)
to the ``primaries'' ${\tilde \chi}_{j(n,k)}$. It is intriguing that
two-dimensional gravity, believed to be related to \G\ Chern-Simons
theory on completely different grounds \hverlinde , exhibits
a similar property \ref\lz{B. Lian and G. Zuckerman, \pl\ {\bf B254}
(1991) 417.}.

\beginsection 3. VERLINDE ALGEBRAS

\indent The algebra of the observables of Chern-Simons theory with compact
gauge group is the Verlinde algebra \verfusion\ of the underlying
two-dimensional conformal field theory, and the maximally commuting
subalgebra of the Chern-Simons observables is the fusion algebra of the
conformal theory.  It is interesting, therefore, to construct
for the \G\ Chern-Simons theory the corresponding objects
whose two-dimensional counterparts are not yet understood.
{}From the previous discussion it follows that the
\G\ Chern-Simons algebra of observables is the image of $C(\tHp)$,
the $C$-invariant subalgebra of the 't Hooft algebra $\tHp$,
in the representations \kac -\kactris .

The Verlinde basis of an algebra of observables might be defined as
a basis $\{\Phi_n(a), \Phi_n(b)\}$ with the properties
\eqn\propertys{\eqalign{\Phi_n(b) &= S^{-1} \Phi_n(a) S \cr
\Phi_n(a)\Phi_m(a) &= \sum_k N_{nm}^k\Phi_k(a) \cr
\Phi_0 (a)&=\Phi_0(b)=Id, \cr}}
where $N_{nm}^k$ are positive integers. Moreover, there should be
a basis $\{v_m \}$ of eigenvectors of $\vopa$,
$$\Phi_n(a) v_m = \lambda_n^{(m)} v_m,$$
\noindent satisfying the equation:
$$\{v_n=\vopb v_0\}.$$
The existence of such a basis is not guaranteed in general, but appears
to be a specific property of Chern-Simons observables algebras,
like $\tHp$ and its $C$-invariant subalgebras.
The Verlinde basis for $\tHp$ is well known:
$$\{\Phi_n(a)=A^n, \Phi_n(b)=B^n, \;\;n=0,1,...,p-1\}.$$

A basis for $C(\tHp)$ is obviously
$$\{\Phi_n(a)=A^n +A^{-n}, \Phi_n(b)=B^n+B^{-n}, \;\;n=0,1,...,[p/2]\},$$
\noindent where $[p/2]$ is the integer part of $p/2$. This basis
is {\it not}, however, a Verlinde basis since the associated
fusion rules $N_{nm}^k$ may be negative. Moreover, the image
of $C(\tHp)$ in the representations \kac -\kactris\ is
$[{p-1\over 2}]$-dimensional if $pq$ is even, but
${p+1\over 2}$-dimensional if $pq$ is odd.
To obtain the Verlinde basis,
it is useful to recall Verlinde's observation \verfusion\ that the eigenvalues
$\lambda_n^{(m)}$ of $\vopa$ are related to the modular matrix $S$ in the
basis $\{v_n\}$ through the equation
\eqn\verfor{\eigen = {S_{mn}\over S_{m0}}.}

When $p$ is even, the holomorphic basis of $\HCS$ in Eq.\kac\
defines a Verlinde basis if one takes $v_m = |m+1>$, with $m=0,1,...,p/2-2$.
In this basis, the modular matrix is
\eqn\smatrix{S_{mn}= {2\over \sqrt{p}}\sin 2\pi{q(n+1)(m+1)\over p},}
\noindent and thus:
\eqn\diago{\vopa |M> = {\sin {2\pi q(n+1)M\over p}\over {\sin 2\pi q M
\over p}}|M>.}
Defining $A = e^{i\theta}$ and $x= 2\cos \theta$,  we obtain
\eqn\cheby{\vopa = {\sin (n+1)\theta \over \sin \theta} =
P_n (x), }
where $P_n (x)$ are the Chebyshev polynomials of the second kind.
It follows from $\cos \theta|M> = \cos {2\pi qM\over p}|M>$, that
\eqn\erelation{ P_{p/2-1}(x) =0.}
Therefore, for $p$ even, the Verlinde algebra is the same
as the Verlinde algebra of $SU(2)$ current algebra of level
$p/2-2$ \ref\gepner{D. Gepner,\cmp\ {\bf 141} (1991) 381.}
and is, in fact, independent of the (odd) number $q$.

If $p$ is odd and $q$ even, the basis defined above is not of Verlinde
type since it leads to fusion rules which are not all positive.
However, the basis
$$v_m = |{p-1\over 2}-m>\;\;\; m=0,1,...,{p-3\over 2}$$
\noindent is of Verlinde type, since from the Verlinde formula
one obtains the equation
\eqn\chebyodd{\vopa = {\sin ({p-1\over 2} - n)\theta \over
\sin {p-1\over 2}\theta} = {\sin (2n+1)\theta /2 \over \sin \theta /2}
= P_{2n} (y),}
where $y = 2\cos \theta /2$. Again, Eq.\representation\ implies the
relation:
\eqn\orelation{ P_{p-1}(y)=0.}
The fusion algebra for $p$ odd is therefore independent of the (even)
number $q$:
\eqn\oddfusion{\Phi_n(a)\Phi_m(a) = \sum_{k=|m-n|}^{min\{m+n; p-2-m-n\}}
\Phi_k(a).}

Finally, let us consider the case when both $p$ and $q$ are odd.
Because of the relations \kactris , we expect to obtain in this
case the fusion algebra of the $D_{p+1}$ diagonal series.
Let us show that the basis
$v_m = |{p-1\over 2}-m>$, with $m=0,1,...,{p-1\over 2}\equiv \nu$
is of Verlinde type. From the expression for the $S$ matrix
in this basis we get the eigenvalues $\eigen$ by means of
the Verlinde formula:
\eqn\dauto{\eigen = (-1)^n{\cos {2\pi q\over p} (\nu -n)(\nu -m)\over
\cos {2\pi q\over p}\nu (\nu-m)}.}
Defining $A= -e^{i\ph}$, the eigenvalues of $\cos n\ph$ are
$\cos{2\pi qnm\over p}$, as follows from the representation
\representation . Thus, the Verlinde operators are:
\eqn\chebyoo{\vopa = (-1)^n{\cos (\nu -n)\ph \over \cos \nu \ph}=
(-1)^n {\cos (2n+1)\ph /2 \over \cos \ph /2}.}
By introducing the variable $z= -2\sin \ph /2$, one can rewrite
the Verlinde operators once again in terms of Chebyshev polynomials
\eqn\bischeby{\vopa = P_{2n}(z) \;\;\; n=0,1,...,\nu .}
However, the fusion algebra is {\it not} the same as for $q$ even
(Eq.\oddfusion) since the generator $P_{2\nu}(z)$
does not vanish on the $({p+1\over 2})$-dimensional
representation space $\HCS$. The polynomial relation
among the $P_{2n}(z)$'s can be easily derived from Eq.\chebyoo :
\eqn\oorelation{ P_{2\nu +2}(z)-P_{2\nu-2}(z)=0.}
The ring defined by Eqs.\bischeby\ and \oorelation\ is indeed the
fusion ring of the $D_{2\nu +2}=D_{p+1}$ series
\ref\dfzu{P. Di Francesco and J.-B. Zuber, Saclay preprint 92/138,
hep-th/9211138.}, with $\Phi_{\nu}(a)=
\Phi_{\nu}^{(+)}+\Phi_{\nu}^{(-)}$ being the sum of the two
``degenerate'' blocks $\Phi_{\nu}^{(\pm)}$ of the $D$ models,
which cannot be distinguished, of course, by the Chern-Simons
theory on the torus.

Having produced the Verlinde representation of the algebra
of Chern-Simons observables $C(\tHp)$, one can exploit the
't Hooft algebras isomorphism of Eq.\decomposition\ to derive explicit
polynomial representations of the fusion rings of the Virasoro
minimal models and of the Kac-Wakimoto set of \G\ current algebra
representations.

The minimal model $(r,s)$, with $r$ {\it odd}, is realized by
taking the $C(\tHq)\times C(\tHp)$ subalgebra of the 't Hooft
algebra $\tHpq$ acting on the $C_{q/p}$ and $C_{p/q}$ odd
subrepresentations of the representation \holoeven , with
$p=2s$ and $q=r$ \cimbimbo . It follows that the fusion algebra
of the minimal models is given by the products
$$P_M(x) P_{2N}(y)$$
\noindent of the Verlinde algebras \cheby\ and \chebyodd\
with $M=0,1,...,s-2$ and $N=0,1,...,{r-3\over 2}$. The
relations \erelation ,\orelation\ become
\eqn\mmrelation{ P_{s-1}(x)= P_{r-1}(y)=0.}
The identification with the standard primary fields of the minimal
models $\phi_{(m,n)}$ labelled by the Kac indices $m,n$ ($1\leq m\leq s-1,
1\leq n\leq {r-1\over 2}$) is obtained by comparing the Rocha-Caridi
characters with the Chern-Simons holomorphic wave functions
\kac . One obtains
\eqn\identification{\phi_{(m,n)} =\cases{P_{m-1}(x) P_{n-1}(y) &if
$n$ odd \cr P_{s-1-m}(x) P_{r-1-n}(y) &if $n$ even\cr}.}

The variables $x= 2\cos\theta$ and $y= 2\cos {\tilde \theta}/2$ are not
independent, since the operators $A=e^{i\theta}$ and ${\tilde A}
=e^{i{\tilde \theta}}$ are both related to the same operator
${\hat A}= e^{i{\hat \theta}}$ of the ``parent'' 't Hooft algebra
$\tHpq$. Eq. \iso\ implies
\eqn\xyrel{x=2\cos r{\hat \theta},\;\;\;  y=2\cos s{\hat \theta},}
from which one derives the identity
\eqn\xypol{P_{r-2}(y)=P_{s-2}(x).}

This equation can be solved to eliminate $y$ from Eq.\identification\
and to write a representation of the minimal models fusion algebra
in terms of one-variable polynomials \dfzu .
Substituting $y=2\cos s{\hat \theta}$ in Eq.\identification\ with
one of its eigenvalues, $y|1> = 2\cos {\pi \over r}|1>\equiv
\gamma |1>$, one reproduces the representation found in \dfzu :
\eqn\xiden{\phi_{(m,n)} =\cases{P_{m-1}(x) P_{n-1}(\gamma) &if
$n$ odd \cr P_{s-1-m}(x) P_{r-1-n}(\gamma)=\cr
P_{s-1-m}(x) P_{n-1}(\gamma) &if $n$ even\cr}.}

The result of \dfzu , that the fusion rings of the $D$ series
and of the minimal models can be represented in terms of one-variable
polynomials (which are essentially Chebyshev
polynomials), can be understood in the Chern-Simons
framwork as consequence of the fact that
all of these models, together with the $A$ series models,
are contained in the ``parent'' 't Hooft algebra $\tHpq$.

A polynomial representation of the fusion algebra defined by
the Kac-Wakimoto representations is obtained in a similar
way:
\eqn\kwrepr{\chi_{j(n,k)} =(-1)^k\cases{P_{n-1}(x) y^{k\over 2}&if
$k$ even \cr P_{2u+t-1-n}(x) y^{k+u\over 2} &if $k$ odd\cr},}
with $n=1,...,2u+t-1$, $k=0,1,...,u-1$, and $u$ odd. A similar
expression can be found for $u$ even. The polynomial relations defining
the fusion ring are
\eqn\kwrelation{P_{2u+t-1}(x)=0, \;\; y^u-1=0.}
It is easy to check that the Kac-Wakimoto operators
${\tilde \chi}_{j(n,k)}= (-1)^k\chi_{j(n,k)}$ do indeed generate
a fusion algebra with positive fusion rules:
\eqn\kwfusion{{\tilde \chi}_{j(n,k)}{\tilde \chi}_{j(n',k')} =
\sum_{m=|n-n'|+1}^{min\{n+n'-1; 4u+2t-1-n-n'\}}\cases{{\tilde \chi}_{j(m,k+k')}
&if $k+k'<u$\cr{\tilde \chi}_{j(2u+t-m,k+k'-u)} &if $k+k'\geq u$.\cr}}

\beginsection 4. CONCLUSIONS

\indent \G\ Chern-Simons theory with fractional charge motivates the study
of the algebra $C(\tHp)$ whose representation theory gives rise
to very simple and natural generalizations of the modular representations
and fusion rings of the $SU(2)$ WZW models. And yet, the two-dimensional
interpretation of these algebraic data appears to lie outside
standard conformal field theory. It is intriguing, in particular,
that the Kac-Wakimoto characters, which cannot have a conformal
field theory interpretation because they have non-positive
fusion rules, do appear in \G\ Chern-Simons theory, though
their relation to Chern-Simons states is of a novel type and
they are multiplied by exactly those minus signs which make their
fusions positive.

\beginsection ACKNOWLEDGEMENTS

\indent I would like to thank A. Schwimmer for bringing reference \dfzu\
to my attention.
\listrefs
\bye